# METHOD OF FORMING STABLE STATES OF DENSE HIGH-TEMPERFATURE PLASMA

## S.I. Fisenko, I.S. Fisenko

"Rusthermosinthes" JSC,
6 Gasheka Str., 12<sup>th</sup> Storey,
Moscow 125047
10 Phone: 956-82-46, Fax: 956-82-47
E-mail: StanislavFisenko@yandex.ru

## ABSTRACT

15      A method is proposed for forming stable states of a dense high-temperature plasma, including plasmas for controlled fusion, the method comprising:  generating a dense high-temperature plasma in pulsed heavy-current discharges, followed by injecting the plasma from the area of a magnetic field with parameters corresponding to the conditions of gravitational emission of electrons with a banded energy spectrum    and
20 subsequent energy transfer along the spectrum (cascade transition) into the long wavelength region (of eV-energy), this leading to the state of locking and amplification of the gravitational emission in the plasma with simultaneous compression thereof to  the states of hydrostatic equilibrium, with using multielectron atoms as a prerequisite element in the composition of a working gas, for quenching the spontaneous gravitational emission
25 from the ground energy levels (the keV-region) of the electron in the proper gravitational field.

PACS 1999 – 04.90.+e, 52.55.Ez, 23.40.-s.

Keywords: gravity, electron, spectrum, discharge, fusion

30

## State of the Art

The existing state of the art related to the realization of stable states of a dense high-temperature plasma applicable for the purposes of nuclear fusion can be defined as a stage of the formation and confinement of a plasma by a magnetic field in devices which
35 make it possible to realize separate techniques of the claimed method but not the method as such, i.e., a method of achieving a stable state of a dense high-temperature plasma. In this respect the claimed method has no close analogs.

From the state of the art a heavy-current pulsed discharge is known, which is shaped with the aid of a cylindrical discharge chamber (whose end faces function as electrodes) which is filled with a working gas (deuterium, hydrogen, a deuterium-tritium mixture at a pressure of 0.5 to10 mm Hg, or noble gases at a pressure of 0.01 to0.1 mm Hg). Then a discharge of a powerful capacitor battery is effected through the gas, with the voltage of 20 to 40 kV supplied to the anode and the current in the forming discharge reaching about 1 MA. In experiments (Lukyanov S.Yu. "Hot Plasma and Controlled Fusion", Moscow, Atomizdat, 1975 (in Russian)) first a first phase of the process was observed — plasma compression to the axis by the current magnetic field with decrease of the current channel diameter by approximately a factor of 10 and formation a brightly glowing plasma column on the discharge axis (z-pinch). In the second phase of the process a rapid development of current channel instabilities (kinks, helical disturbances, etc.) were observed.

The buildup of these instabilities occurs very rapidly and leads to the degradation of the plasma column (plasma jets outburst, discharge discontinuities, etc.), so that the discharge lifetime is limited to a value on the order of $10'''$ s. For this reason in a linear pinch it turns out to be unreal to fulfill the conditions of nuclear fusion defined by the Lawson criterion $n\tau > 10^{14}$ cm$^{-3}$s, where n is the plasma concentration, $\tau$ is the discharge lifetime..

A similar situation takes place in a $\Theta$-pinch, when to a cylindrical discharge chamber an external longitudinal magnetic field inducing an azimuthal current is impressed.

Magnetic traps are known, which are capable of confining a high-temperature plasma for a long time (but not sufficient for nuclear fusion to proceed) within a limited volume (see Artsimovich L.A., "Closed Plasma Configurations", Moscow, Atomizdat, 1969 (in Russian)). There exist two main varieties of magnetic traps: closed and open ones.

Magnetic traps are devices which are capable of confining a high-temperature plasma for a sufficiently long time within a limited volume and which are described in Artsimovich L.A., "Closed Plasma Configurations", Moscow, Atomizdat, 1969.

To magnetic traps of closed type (on which hopes to realize the conditions of controlled nuclear fusion (CNU) were pinned for a long time) there belong devices of the Tokamak, Spheromak and Stellarator type in various modifications (Lukyanov S.Yu., "Hot Plasma and CNU", Moscow, Atomizdat, 1975 (in Russian)).

5      In devices of the Tokamak type a ring current creating a rotary transformation of magnetic lines of force is excited in the very plasma. Spheromak represents a compact torus with a toroidal magnetic field inside a plasma. Rotary transformation of magnetic lines of force, effected without exciting a toroidal current in plasma, is realized in Stellarators (Volkov E.D. et al., "Stellarator", Moscow, Nauka, 1983 (in Russian)).

10      Open-type magnetic traps with a linear geometry are: a magnetic bottle, an ambipolar trap, a gas-dynamic type trap (Ryutov D.D., "Open traps", Uspekhi Fizicheskikh Nauk, 1988, vol. 154, p. 565).

In spite of all the design differences of the open-type and closed-type traps, they are based on one principle: attaining hydrostatic equilibrium states of plasma in a magnetic field through the equality of the gas-kinetic plasma pressure and of the magnetic field pressure at the external boundary of plasma. The very diversity of these traps stems from the absence of positive results.

When using a plasma focus device (**PF**) (this is how an electric discharge is called), a non-stationary bunch of a dense high-temperature (as a rule, deuterium) plasma is obtained (this bunch is also called "plasma focus"). PF belongs to the category of pinches and is formed in the area of current sheath cumulation on the axis of a discharge chamber having a special design. As a result, in contradistinction to a direct pinch, plasma focus acquires a non-cylindrical shape (Petrov D.P. et al., "Powerful pulsed gas discharge in chambers with conducting walls" in Collection of Papers "Plasma Physics and Problem of Controlled Thermonuclear Reactions", volume 4, Moscow, Izdatel'stvo AN SSSR, 1958 (in Russian)).

Unlike linear pinch devices, where the function of electrodes is performed by the chamber end faces, in the PF the role of the cathode is played by the chamber body, as a result of which the plasma bunch acquires the form of a funnel (thence the name of the device). With the same working parameters as in the cylindrical pinch, in a PF device a plasma having higher temperature, density and longer lifetime is obtainable, but the subsequent development of the instability destroys the discharge, as is the case in the linear pinch (Burtsev B.A. et al., "High-temperature plasma formations" in: Itogi Nauki i

Tekhniki", "Plasma Physics" Series, vol. 2, Moscow, Izdatel'stvo AN SSSR, 1981 (in Russian)), and stable states of plasma are actually not attained.

Non-stationary bunches of high-temperature plasma are also obtained in gas-discharge chambers with a coaxial arrangement of electrodes (using devices with coaxial plasma injectors). The first device of such kind was commissioned in 1961 by J. Mather (Mather J.W., "Formation on the high-density deuterium plasma focus", Phys. Fluids, 1965, vol. 8, p. 366). This device was developed further (in particular, see (J.Brzosko et al., Phys. Let. A., 192 (1994), p. 250, Phys. Let. A., 155 (1991), p. 162)). An essential element of this development was the use of a working gas doped with multielectron atoms. Injection of plasma in such devices is attained owing to the coaxial arrangement of cylindrical chambers, wherein the internal chamber functioning as the anode is disposed geometrically lower than the external cylinder — the cathode. In the works of J.Brzosko it was pointed out that the efficiency of the generation of plasma bunches increases when hydrogen is doped with multielectron atoms. However, in these devices the development of the instability substantially limits the plasma lifetime as well. As a result, this lifetime is smaller than necessary for attaining the conditions for a stable course of the nuclear fusion reaction. With definite design features, in particular, with the use of conical coaxial electrodes (M.P. Kozlov and A.I. Morozov (Eds.), "Plasma Accelerators and Ion Guns", Moscow, Nauka, 1984 (in Russian)), such devices are already plasma injection devices. In the above-indicated devices (devices with coaxial cylindrical electrodes) plasma in all the stages up to the plasma decay, remains in the magnetic field area, though injection of plasma into the interelectrode space takes place. In pure form injection of plasma from the interelectrode space is observed in devices with conical coaxial electrodes. The field of application of plasma injectors is regarded to be auxiliary for plasma injection with subsequent use thereof (for example,. for additional pumping of power in devices of Tokamak type, in laser devices, etc.), which, in turn, has limited the use of these devices not in the pulsed mode, but in the quasi-stationary mode.

Thus, the existing state of the art, based on plasma confinement by a magnetic field, does not solve the problem of confining a dense high-temperature plasma during a period of time required for nuclear fusion reactions to proceed, but effectively solves the problem of heating plasma to a state in which these reactions can proceed.

## Disclosure of the Invention

The author proposes a solution of the above-stated problem, which can be attained by a new combination of means (devices) known in the art with the use of their combination (the parameters considered earlier), which was not only not used heretofore, but proposed or supposed in the state of the art, and which is further described in detail in the sections dealing with carrying the invention into effect and in the set of claims.

Accordingly, the present invention relates to a method of forming stable states of a dense high-temperature plasma, which comprises the following steps:

a) generation of a dense high-temperature plasma from hydrogen and isotopes thereof with the aid of pulsed heavy-current discharges;

b) injection of the plasma from the area of a magnetic field with parameters corresponding to the conditions of gravitational emission of electrons with a banded energy spectrum;

c) energy transfer along the spectrum.

The energy transfer (step c) is performed by cascade transition into the long wavelength region of eV-energy to the state of locking and amplification of the gravitational emission and simultaneous compression to the states of hydrostatic equilibrium, and in the formation of said states in the composition of a working gas multielectron atoms are used for quenching the spontaneous gravitational emission from the ground energy levels of the keV-region electron in the proper gravitational field.

It is preferable that in one of the embodiment of the invention for obtaining stable states of a dense high-temperature plasma use is made of hydrogen and multielectron atoms, such as krypton, xenon, and other allied elements (neon, argon).

In another preferable embodiment of the invention in order to realize the conditions for the nuclear fusion reaction to proceed use is made of hydrogen and carbon, wherein carbon is also employed both for quenching the spectra of gravitational emission with keV energies and as a fusion reaction catalyst.

The claimed method provides a scheme for forming stable states of a dense high-temperature plasma, which scheme comprises a device for supplying a working gas, a discharge chamber, a discharge circuit, a chamber for forming a stable plasma bunch.

If and when necessary, each of the cited blocks of the scheme can be fitted with appropriate measuring equipment.

The invention is illustrated by a circuit diagram of a pulsed heavy-current magnetic-compression discharge on multiply charged ions with conical coaxial electrodes, in which :

1. a fast-acting valve for supplying a working gas into a gap between an internal electrode (2) and an external electrode (3);

2. an external electrode;

3. an internal electrode has a narrowing surface close to conical one;

4. a diverter channel which prevents the entrance of admixtures into the compression area;

5. a discharge circuit;

6. an area of compression by a magnetic field;

7. an area of compression due to efflux current in the outgoing plasma jet and subsequent compression by the emitted gravitational field.

## Carrying out the Invention

## Terms and Definitions Used in the Application

The definition *"stable states of a dense high-temperature plasma"* denotes the states of hydrostatic equilibrium of a plasma, when the gas-dynamic pressure is counterbalanced by the pressure of a magnetic field or, in the present case, by the pressure of the emitted gravitational field.

The definition *"a dense high-temperature plasma"* denotes a plasma with the lower values of densities $n_C$, $n_i = (10^{23} - 10^{25})$ m$^{-3}$ and temperatures $T_c$, $T_i = (10^7 - 10^8)$ K.

The definition *"plasma parameters corresponding to gravitational emission of electrons"* (with a banded emission spectrum) denotes parameters which are in the above-indicated range of pressures and temperatures.

The definition *"locking of gravitational emission in plasma"* denotes the state of gravitational emission in a plasma ,which takes place when its emission frequency and electron Langmuir frequency are equal. In the present case locking of the emission takes place for two reasons:

- energy transfer along the spectrum into the long wavelength region as a result of cascade transitions into the long wavelength region with attaining emission frequency $(10^{13} - 10^{14})$ with plasma Langmuir frequency equal to the electron one, this being the condition of locking gravitational emission in plasma;

- quenching spontaneous gravitational emission of electrons from the ground energy levels by multielectron atoms, when the energy of an excited electron is transferred to an ion with corresponding energy levels, leading to its ionization.

The definition *"amplification of gravitational emission"* denotes amplification which takes place when the gravitational emission is locked, because, with the locking conditions having been fulfilled, the gravitational emission remains in plasma with sequential emission of the total energy of the gravitational field locked in the plasma.

For a better understanding of the invention, given below is a description of high-temperature plasma formations which take place in the proposed method, and a description of a method of forming their stable states as hydrostatic equilibrium states. The conditions of gravitational emission of electrons with a banded spectrum, the conditions of exciting gravitational emission in plasma, and locking and amplification owing to cascade transitions as claimed in the set of claims presented hereinbelow, are disclosed.

**1. Gravitational emission of electrons with a banded spectrum as emission of the same level with electromagnetic emission.**

For a mathematical model of interest, which describes a banded spectrum of stationary states of electrons in the proper gravitational field, two aspects are of importance. First. In Einstein's field equations $\kappa$ is a constant which relates the space-time geometrical properties with the distribution of physical matter, so that the origin of the equations is not connected with the numerical limitation of the $\kappa$ value. Only the requirement of conformity with the Newtonian Classical Theory of Gravity leads to the small value $\kappa = 8\pi G/c^4$, where $G$, $c$ are, respectively, the Newtonian gravitational constant and the velocity of light. Such requirement follows from the primary concept of the Einstein General Theory of Relativity (**GTR**) as a relativistic generalization of the Newtonian Theory of Gravity. Second. The most general form of relativistic gravitation equations are equations with the $\Lambda$ term. The limiting transition to weak fields leads to the equation

$$\Delta\Phi = -4\pi\rho G + \Lambda c^2,$$

where $\Phi$ is the field scalar potential, $\rho$ is the source density. This circumstance, eventually, is crucial for neglecting the $\Lambda$ term, because only in this case the GTR is a generalization of the Classical Theory of Gravity. Therefore, the numerical values of $\kappa = $ $8\pi G/c^4$ and $\Lambda = 0$ in the GTR equations are not associated with the origin of the equations, but follow only from the conformity of the GTR with the classical theory.

From the 70's onwards, it became obvious (Siravam C . and Sinha K., Phys. Rep. 51 (1979) 112) that in the quantum region the numerical value of $G$ is not compatible with the principles of quantum mechanics. In a number of papers (Siravam C . and Sinha K., Phys. Rep. 51 (1979) 112) (including also Fisenko S. et al., Phys. Lett. A, 148,,7,9 (1990) 405)) it was shown that in the quantum region the coupling constant $K$ ($K \approx 10^{40}$ $G$) is acceptable. The essence of the problem of the generalization of relativistic equations on the quantum level was thus outlined: such generalization must match the numerical values of the gravity constants in the quantum and classical regions.

In the development of these results, as a micro-level approximation of Einstein's field equations, a model is proposed, based on the following assumption:

"*The gravitational field within the region of localization of an elementary particle having a mass $m_0$ is characterized by the values of the gravity constant $K$ and of the constant $\Lambda$ that lead to the stationary states of the particle in its proper gravitational field, and the particle stationary states as such are the sources of the gravitational field with the Newtonian gravity constant $G$*".

The most general approach in the Gravity Theory is the one which takes twisting into account and treats the gravitational field as a gage field, acting on equal terms with other fundamental fields (Ivanenko et al., Gage Theory of Gravitation, Moscow, MGU Publishing House, 1985 (in Russian)). Such approach lacks in apriority gives no restrictions on the microscopic level. For an elementary spinor source with a mass $m_0$, the set of equations describing its states in the proper gravitational field in accordance with the adopted assumption will have the form

$$\left\{ i\gamma^\mu \left( \nabla_\mu + \overline{\kappa} \Psi \gamma_\mu \gamma_5 \Psi \gamma_5 \right) - m_0 c/\hbar \right\} \Psi = 0 \tag{1}$$

$$R_{\mu\nu} - \frac{1}{2} g_{\mu\nu} R = -\kappa \left\{ T_{\mu\nu}\left( E_n \right) - \mu g_{\mu\nu} + \left( g_{\mu\nu} S_\alpha S^\alpha - S_\mu S_\nu \right) \right\} \tag{2}$$

$$R\left( K, \Lambda, E_n, r_n \right) = R\left( G, E_n', r_n \right) \tag{3}$$

$$\left\{ i\gamma^\mu \nabla_\mu - m_n c/\hbar \right\} \Psi' = 0 \tag{4}$$

$$R_{\mu\nu} - \frac{1}{2} g_{\mu\nu} R = -\kappa' T_{\mu\nu}\left( E_n' \right) \tag{5}$$

The following notations are used throughout the text of the paper: $\kappa = 8\pi K/c^4$, $\kappa' = 8\pi G/c^4$, $E_n$ is the energy of stationary states in the proper gravitational field with the constant $K$, $\Lambda = \kappa\mu$, $r_n$ is the value of the coordinate $r$ satisfying the equilibrium n-state in the proper gravitational field, $\overline{\kappa} = \kappa_0\kappa$, $\kappa_0$ is the dimensionality constant, $S_a = \overline{\Psi}\gamma_a\gamma_5\Psi$,

5    $\nabla_\mu$ is the spinor-coupling covariant derivative independent of twisting, $E'_n$ is the energy state of the particle having a mass $m_n$ (either free of field or in the external field), described by the wave function $\psi'$ in the proper gravitational field with the constant $G$. The rest of the notations are generally adopted in the gravitation theory.

Equations (1) through (5) describe the equilibrium states of particles (stationary

10    states) in the proper gravitational field and define the localization region of the field characterized by the constant $K$ that satisfies the equilibrium state. These stationary states are sources of the field with the constant $G$, and condition (3) provides matching the solution with the gravitational constants $K$ and $G$. The proposed model in the physical aspect is compatible with the principles of quantum mechanics principles, and the

15    gravitational field with the constants $K$ and $\Lambda$ at a certain, quite definite distance specified by the equilibrium state transforms into the filed having the constant $G$ and satisfying, in the weak field limit, the Poisson equation.

The set of equations (1) through (5), first of all, is of interest for the problem of stationary states, i.e., the problem of energy spectrum calculations for an elementary

20    source in the own gravitational field. In this sense it is reasonable to use an analogy with electrodynamics, in particular, with the problem of electron stationary states in the Coulomb field. Transition from the Schrödinger equation to the Klein-Gordon relativistic equations allows taking into account the fine structure of the electron energy spectrum in the Coulomb field, whereas transition to the Dirac equation allows taking into account the

25    relativistic fine structure and the energy level splitting associated with spin-orbital interaction. Using this analogy and the form of equation (1), one can conclude that solution of this equation without the term $\overline{\kappa}\overline{\Psi}\gamma_\mu\gamma_5\Psi\gamma_5$ gives a spectrum similar to that of the fine structure (similar in the sense of relativism and removal of the principal quantum number degeneracy).. Taking the term $\overline{\kappa}\overline{\Psi}\gamma_\mu\gamma_5\Psi\gamma_5$ into account, as is noted in Siravam

30    C. and Sinha K., Phys. Res. 51 (1979) 112 , is similar to taking into account of the term $\overline{\Psi}O^{\mu\nu}\Psi F_{\mu\nu}$ in the Pauli equation. The latter implies that the solution of the problem of stationary states with twisting taken into account will give a total energy-state spectrum with both the relativistic fine structure and energy state splitting caused by spin-twist interaction taken into account. This fact, being in complete accord with the requirements of the Gauge Theory of Gravity, allows us to believe that the above-stated assumptions concern ing the properties of the gravitational field in the quantum region are relevant, in the general case, just to the gravitational field with twists.

Complexity of solving this problem compels us to employ a simpler approximation, namely,: energy spectrum calculations in a relativistic fine-structure approximation. In this approximation the problem of the stationary states of an elementary source in the proper gravitational field well be reduced to solving the following equations:

$$f'' + \left(\frac{\nu' - \lambda'}{2} + \frac{2}{r}\right)f' + e^{\lambda}\left(K_n^2 e^{-\nu} - K_0^2 - \frac{l(l+1)}{r^2}\right)f = 0 \tag{6}$$

$$-e^{-\lambda}\left(\frac{1}{r^2} - \frac{\lambda'}{r}\right) + \frac{1}{r^2} + \Lambda = \beta(2l+1)\left\{f^2\left[e^{-\lambda}K_n^2 + K_0^2 + \frac{l(l+1)}{r^2}\right] + f'^2 e^{-\lambda}\right\} \tag{7}$$

$$-e^{-\lambda}\left(\frac{1}{r^2} - \frac{\nu'}{r}\right) + \frac{1}{r^2} + \Lambda = \beta(2l+1)\left\{f^2\left[K_0^2 - K_n^2 e^{-\nu} + \frac{l(l+1)}{r^2}\right] - e^{\lambda}f'^2\right\} \tag{8}$$

$$\left\{-\frac{1}{2}(\nu'' + \nu'^2) - (\nu' + \lambda')\left(\frac{\nu'}{4} + \frac{1}{r}\right) + \frac{1}{r^2}(1 + e^{\lambda})\right\}_{r=r_n} = 0 \tag{9}$$

$$f(0) = const << \infty \tag{10}$$

$$f(r_n) = 0 \tag{11}$$

$$\lambda(0) = \nu(0) = 0 \tag{12}$$

$$\int_0^{r_n} f^2 r^2 dr = 1 \tag{13}$$

Equations (6)—{8} follow from equations (14)—( 15)

$$\left\{-g^{\mu\nu}\frac{\partial}{\partial x_\mu}\frac{\partial}{\partial x_\nu} + g^{\mu\nu}\Gamma_{\mu\nu}^{\alpha}\frac{\partial}{\partial x_\alpha} - K_0^2\right\}\Psi = 0 \tag{14}$$

$$R_{\mu\nu} - \frac{1}{2}g_{\mu\nu}R = -\kappa\left(T_{\mu\nu} - \mu g_{\mu\nu}\right), \tag{15}$$

after the substitution  of $\Psi$ in the form: $\Psi = f_{El}(r)Y_{lm}(\theta, \varphi)\exp\left(\frac{-iEt}{\hbar}\right)$  into them and specific computations in the central-symmetry field metric with the interval defined by the expression (Landau L.D., Lifshitz E.M., Field Theory, Moscow, Nauka Publishers, 1976)

$$dS^2 = c^2 e^{\nu} dt^2 - r^2\left(d\theta^2 + \sin^2\theta d\varphi^2\right) - e^{\lambda}dr^2 \tag{16}$$

The following notation is used above: $f_m$ is the radial wave function that describes the states with a definite energy $E$ and the orbital moment $l$ (hereafter the subscripts $El$ are omitted), $Y_{lm}(\theta, \varphi)$ - are spherical functions, $K_n = E_n /\hbar c$, $K_0 = cm_0 /\hbar$, $\beta = (\kappa/4\pi)(\hbar/m_0)$.

Condition (9) defines $r_n$, whereas equations (10) through (12) are the boundary conditions and the normalization condition for the function $f$, respectively. Condition (9) in the general case has the form $R(K,r_n) = R(G,r_n)$. Neglecting the proper gravitational field with the constant $G$, we shall write down this condition as $R(K,r_n) = 0$, to which equality (9) actually corresponds.

The right-hand sides of equations (7)—(8) are calculated basing on the general expression for the energy-momentum tensor of the complex scalar field:

$$T_{\mu\nu} = \Psi_{,\mu}^{+}\Psi_{,\nu} + \Psi_{,\nu}^{+}\Psi_{,\mu} - \left(\Psi_{,\mu}^{+}\Psi^{,\mu} - K_0^2\Psi^{+}\Psi\right) \qquad (17)$$

The appropriate components $T_{\mu\nu}$ are obtained by summation over the index $m$ with application of characteristic identities for spherical functions (Warshalovich D.A. et al., Quantum Theory of Angular Momentum, Leningrad, Nauka Publishers, 1975 (in Russian))

after the substitution of
$$\Psi = f(r)Y_{lm}(\theta, \varphi)\exp\left(\frac{-iEt}{\hbar}\right)$$
into (17).

Even in the simplest approximation the problem of the stationary states of an elementary source in the proper gravitational field is a complicated mathematical problem. It becomes simpler if we confine ourselves to estimating only the energy spectrum. Equation (6) can be reduced in many ways to the equations (E. Kamke, Differentialgleichungen, Lösungsmethoden und Lösungen, Leizig, 1959)

$$f' = fP(r) + Q(r)z \qquad\qquad z' = fF(r) + S(r)z \qquad (18)$$

This transition implies specific choice of $P$, $Q$, $F$, $S$, such that the conditions

$$P + S + Q'/Q + g = 0 \qquad\qquad FQ + P' + P^2 + Pg + h = 0 \qquad (19)$$

should be fulfilled, where $g$ and $h$ correspond to equation. (6) written in the form: $f'' + gf' + hf = Q$. Conditions (19) are satisfied, in particular, by $P$, $Q$, $F$, $S$ written as follows:

$$Q = 1, \qquad P = S = -g/2, \qquad F = \frac{1}{2}g' + \frac{1}{4}g^2 - h \qquad (20)$$

Solutions of set (18) will be the functions: (E. Kamke, Differentialgleichungen, Lösungsmethoden und Lösungen, Leizig, 1959)

$$f = C\rho(r)\sin\theta(r) \qquad\qquad z = C\rho(r)\cos\theta(r) \qquad (21)$$

where $C$ is an arbitrary constant, $\theta(r)$ is the solution of the equation:

$$\theta' = Q\cos^2\theta + (P - S)\sin\theta\cos\theta - F\sin^2\theta, \tag{22}$$

and $\rho(r)$ is found from the formula

$$\rho(r) = \exp\int_0^r\left[P\sin^2\theta + (Q + F)\sin\theta\cos\theta + S\cos^2\theta\right]dr. \tag{23}$$

In this case, the form of presentation of the solution in polar coordinates makes it possible to determine zeros of the functions $f(r)$ at $r = r_n$, with corresponding values of $\theta = n\pi$ ($n$ being an integer). As one of the simplest approximations for $\nu, \lambda$, we shall choose the dependence:

$$e^\nu = e^{-\lambda} = 1 - \frac{\tilde{r}_n}{r + C_1} + \Lambda(r - C_2)^2 + C_3 r \tag{24}$$

where

$$\tilde{r}_n = \frac{2Km_n}{c^2} = \frac{2KE_n}{c^4} = \left(\frac{2K\hbar}{c^3}\right)K_n, \quad C_1 = \frac{\tilde{r}_n}{\Lambda r_n^2}, \quad C_2 = r_n, \quad C_3 = \frac{\tilde{r}_n}{r_n(r_n + C_1)}$$

Earlier the estimate for $K$ was adopted to be $K \approx 1.7 \times 10^{29}$ Nm$^2$kg$^{-2}$. If we assume that the observed value of the electron rest mass $m_1$ is its mass in the ground stationary state in the proper gravitational field, then $m_0 = 4m_1/3$. From dimensionality considerations it follows that energy in the bound state is defined by the expression $\left(\sqrt{Km_0}\right)^2 / r_1 = 0.17 \times 10^6 \times 1.6 \times 10^{-19}$ J, where $r_1$ is the classical electron radius. This leads to the estimate $K \approx 5.1 \times 10^{31}$ Nm$^2$kg$^{-2}$ which is later adopted as the starting one. Discrepancies in the estimates $K$ obtained by different methods are quite admissible, all the more so since their character is not catastrophic. From the condition that $\mu$ is the electron energy density it follows: $\mu = 1.1 \times 10^{30}$ J/m$^3$, $\Lambda = \kappa\mu = 4.4 \times 10^{29}$ m$^{-2}$. From (22) (with the equation for $f(r)$ taken into account) it follows:

$$2\theta' = (1 - \overline{F}) + (1 + \overline{F})\cos 2\theta \approx (1 - \overline{F}), \tag{25}$$

where

$$\overline{F} = \frac{1}{2}\overline{g}' + \frac{1}{4}\overline{g}^2 - \overline{h}, \quad \overline{g} = r_n\left(\frac{2}{r} + \frac{(\nu' - \lambda')}{2}\right), \quad \overline{h} = r_n^2 e^\lambda\left(K_n^2 e^{-\nu} - K_0^2 - \frac{l(l+1)}{r^2}\right).$$

The integration of equation (25) and substitution of $\theta = \pi n$, $r = r_n$ give the relation between $K_n$ and $r_n$:

$$-2\pi m = -\frac{7}{4} - \frac{r_n K_n^2}{\Lambda^2} \sum_{i=1}^{3} \left\{ A_i \left[ \frac{(r_n + \alpha_i)^2}{2} - 2\alpha_i(r_n + \alpha_i) + \frac{\alpha_i^3}{(r_n + \alpha_i)} + 2C_1(r_n + \alpha_i) + \right. \right.$$

$$\left. + 2C_1 \frac{\alpha_i^2}{r_n + \alpha_i} + \frac{C_2^2 \alpha_i}{r_n + \alpha_i} \right] + B_i \left[ (r_n + \alpha_i) + \alpha_i^2 \frac{1}{r_n + \alpha_i} + \frac{2C_1 \alpha_i}{r_n + \alpha_i} - \frac{C_2^2}{r_n + \alpha_i} \right] \right\} +$$

$$+ \frac{K_0^2 r_n}{\Lambda^2} \sum_{i=1}^{3} A_i'(r_n + \alpha_i) + \frac{r_n l(l+1)}{\Lambda} \left[ d_1 r_n - \frac{C_1 d_2}{r_n} + \sum_{i=1}^{3} a_i(r_n + \alpha_i) \right] - \qquad (26)$$

$$- \frac{K_n^2 r_n}{\Lambda^2} \left\{ \sum_{i=1}^{3} \left[ 2\alpha_i^2 A_i - 2\alpha_i B_i - 4C_1 A_i \alpha_i + 2C_1 B_i + C_2^2 A_i + \frac{K_0^2 \Lambda A_i'}{K_n^2}(\alpha_i - C_1) - \right. \right.$$

$$\left. - r_n^2 \Lambda l(l+1) a_i (C_1 - \alpha_i) \right] \ln(r_n + \alpha_i) - r_n \Lambda^{-1} l(l+1)(d_2 + C_1 d_1) \ln r_n \right\}$$

The coefficients entering into equation (26) are coefficients at simple fractions in the expansion of polynomials, required for the integration, wherein $\alpha_i \sim K_n$, $d_2 \sim A_i \sim r_n^{-5}$, $B_i \sim r_n^{-4}$, $A_i' \sim r_n^{-2}$, $a_i \sim r_n^{-4}$, $d_1 = r_n^{-4}$. For eliminating $r_n$ from (26), there exists condition (9) (or the condition exp $v(K,r_n) = 1$ equivalent to it for the approximation employed), but its direct use will complicate the already cumbersome expression (26) still further. At the same time, it easy to note that $r_n \sim 10^{-3} r_{nc}$, where $r_{nc}$ is the Compton wavelength of a particle of the mass $m_n$, and, hence, $r_n \sim 10^{-3} K_n^{-1}$. The relation (26) per se is rather approximate, but, nevertheless, its availability, irrespective of the accuracy of the approximation, implies the existence of an energy spectrum as a consequence of the particle self-interaction with its own gravitational field in the range $r \leq r_n$, where mutually compensating action of the field and the particle takes place. With $l = 0$ the approximate solution (26), with the relation between $r_n$ and $K_n$ taken into account, has the form

$$E_n = E_0 \left( 1 + \alpha e^{-\beta n} \right)^{-1}, \qquad (27)$$

where $\alpha = 1.65$, $\beta = 1.60$.

The relation (27) is concretized, proceeding from the assumption that the observed value of the electron rest mass is the value of its mass in the grounds stationary state in the proper gravitational field, the values $r_1 = 2.82 \times 10^{-15}$ m, $K_l = 0.41 \times 10^{12}$ m$^{-1}$ giving exact zero of the function by the very definition of the numerical values for $K$ and $\Lambda$.

So, the presented numerical estimates for the electron show that within the range of its localization, with $K \sim 10^{31}$ N m$^2$ kg$^{-2}$ and $\Lambda \sim 10^{29}$ m$^{-2}$, there exists the spectrum of stationary states in the proper gravitational field. The numerical value of $K$ is, certainly, universal for any elementary source. *Existence of such numerical value $\Lambda$ denotes a phenomenon having a deep physical sense: introduction into density of the Lagrange function of a constant member independent on a state of the field. This means that the* *time-space has an inherent curving which is connected with neither the matter nor the gravitational waves.* The distance at which the gravitational field with the constant K is localized is less than the Compton wavelength, and for the electron, for example, this value is of the order of its classical radius. At distances larger than this one, the gravitational field is characterized by the constant *G,* i.e., correct transition to Classical GTR holds.

From equation (27) there follow in a rough approximation the numerical values of the stationary state energies: $E_1$ =0.511 MeV, $E_2$ =0.638 MeV, ... $E_\infty$ =0.681 MeV.

Quantum transitions over stationary states must lead to the gravitational emission characterized by the constant *K* with transition energies starting from 127 keV to 170 keV Two circumstances are essential here.

*First.* The correspondence between the electromagnetic and gravitational interaction takes place on replacement of the electric charge e by the gravitational "charge" $m\sqrt{K}$ , so that the numerical values K place the electromagnetic and gravitational emission effects on the same level (for instance, the electromagnetic and gravitational bremsstrahlung cross-sections will differ only by the factor 0.16 in the region of coincidence of the emission spectra).

*Second.* The natural width of the energy levels in the above-indicated spectrum of the electron stationary states will be very small. The small value of the energy level widths, compared to the electron energy spread in real conditions, explains why the gravitational emission effects are not observed as a mass phenomenon in epiphenomena, e.g., in the processes of electron beam bremsstrahlung on targets. However, there is a possibility of registration of gravitational emission spectrum lines. The results are given below. A direct confirmation of the presence of the electron stationary states in the own gravitational field with the constant K may be the presence of the lower boundary of nuclear β-decay. Only starting with this boundary β-asymmetry can take place, which is interpreted as parity non-conservation in weak interactions, but is actually only a consequence of the presence of the excited states of electrons in the own gravitational field in β-decay. Beta-asymmetry was observed experimentally only in β-decay of heavy nuclei in magnetic field (for example, $_{27}C^{60}$ in the known experiment carried out by Wu  (Wu Ts.S., Moshkovskii S.A., Beta-decay, Atomizdat, Moscow,  1970 (in Russian)). On light nuclei, such as $_1H^3$, where the β-decay asymmetry already must not take place, similar experiments were not carried out.

## 2. Conditions of Gravitational Emission in Plasma (Excitation
## of Gravitational Emission in Plasma)

5　　　　For the above-indicated energies of transitions over stationary states in the own field and the energy level widths, the sole object in which gravitational emission can be realized as a mass phenomenon will be, as follows from the estimates given below, a dense high-temperature plasma.

　　　　Using the Born approximation for the bremsstrahlung cross-section, we can write down the expression for the electromagnetic bremsstrahlung per unit of volume per unit of

10 time as

$$\mathbf{Q_e} = \frac{32}{3} \; \frac{z^2 r_0^{\;2}}{137} - mc^2 \; n_e \, n_i \; \frac{\sqrt{2k}\,T_e}{\pi n} = 0.17 \times 10^{-39} \, z^2 n_e n_i \sqrt{T_e}, \qquad \textbf{(28)}$$

where $T_e$, $k$, $n_i$, $n_e$, $m$, $z$, $r_o$ are the electron temperature, Boltzmann's constant, the concentration of the ionic and electronic components, the electron mass, the serial number

15 of the ionic component, the classical electron radius, respectively.

　　　　Replacing $r_o$ by $r_g = 2K\;m/s^2$ (which corresponds to replacing the electric charge e by the gravitational charge $m\sqrt{K}$ ), we can use for the gravitational bremsstrahlung the relation

$$\mathbf{Q_g = 0.16 Q_e.} \qquad \textbf{(29)}$$

20　　　　From (28) it follows that in a dense high-temperature plasma with parameters $n_e = n_i = 10^{23}$ $m^{-3}$, $T_e = 10^7$ K, the specific power of the electromagnetic bremsstrahlung is equal to $\approx 0.53$ $10^{10}$ J/$m^3$ s, and the specific power of the gravitational bremsstrahlung is $0.86$ $10^9$ J/$m^3$ s. These values of the plasma parameters, apparently, can be adopted as guide threshold values of an appreciable gravitational emission level, because the relative

25 proportion of the electrons whose energy on the order of the energy of transitions in the own gravitational field, diminishes in accordance with the Maxwellian distribution exponent as $T_e$ decreases.

## 3. Locking and Amplification of Gravitational Emission by Cascade
30 ## Transitions and Quenching Spontaneous Emission from Ground energy levels
## by Ions of Multielectron Atoms in Plasma Injected from Magnetic Field Area

For the numerical values of the plasma parameters $T_e = T_i = (!0^7 — 10^8)K$, $n_e = n_i = (10^{23} — 10^{25})$ m$^{-3}$ the electromagnetic bremsstrahlung spectrum will not change essentially with Compton scattering of electron emission, and the bremsstrahlung itself is a source of emission losses of a high-temperature plasma. The frequencies of this continuous spectrum are on the order of $(10^{18} — 10^{20})$ s$^{-1}$, while the plasma frequency for the above-cited plasma parameters is $(10^{13} — 10^{14})$ s$^{-1}$, or 0.1 eV of the energy of emitted quanta.

*The fundamental distinction of the gravitational bremsstrahlung from the electromagnetic bremsstrahlung is the banded spectrum of the gravitational emission, corresponding to the spectrum of the electron stationary states in the own gravitational field.*

The presence of cascade transitions from the upper excited levels to the lower ones will lead to that the electrons, becoming excited in the energy region above 100 keV, will be emitted, mainly, in the eV region, i.e., energy transfer along the spectrum to the low-frequency region will take place. Such energy transfer mechanism can take place only in quenching spontaneous emission from the lower electron energy levels in the own gravitational field, which rules out emission with quantum energy in the keV region. A detailed description of the mechanism of energy transfer along the spectrum will hereafter give its precise numerical characteristics. Nevertheless, undoubtedly, the very fact of its existence, conditioned by the banded character of the spectrum of the gravitational bremsstrahlung, can be asserted. The low-frequency character of the gravitational bremsstrahlung spectrum will lead to its amplification in plasma by virtue of the locking condition $\omega_g \leq 0.5\sqrt{10^3 n_e}$ being fulfilled.

It is obvious that these data need to be supplemented with direct experimental identification both regarding both electron gravitational radiation spectrum lines and electron steady state energy spectrum in its own gravitational field. Fig. 2 shows electron beam energy spectra in a pulse accelerator measured by a semicircular magnetic spectrometer. Two peaks of the energy spectra are connected to the feature of the pulse accelerator operations, the secondary pulse is due to lower voltage. This leads to the second (low-energy) maximum of the energy spectrum distribution.

A telemetry error in the middle and soft parts of the spectrum is not more than $\pm$ 2%. The magnetic spectrometer was used for measuring the energy spectrum of electrons after passing through the accelerator anode grid and also spectra of electrons after passing though a foil arranged behind the accelerator mesh anode. These data (and the calculated spectrum) are presented in Fig. 2. Similar measurements were carried out for Ti foil (foil thickness 50 μm) and Ta (foil thickness 10 μm). In case of Ti the measurements were limited from the top by energy of 0.148 MeV, and in case of TA by energy of 0.168 MeV. Above these values the measurement errors increase substantially (for this type of the accelerator). The difference between the normalized spectral densities of theoretical and experimental electron spectra after passing through Ti, Ta and Al foils, Fig. 3. The data indicate that there is a spectrum of electron energy states in their own gravitational field when the electrons are excited when passing through a foil. The obtained data are not sufficient for numerical spectrum identification but the very fact of the spectrum presence according to the data is doubtless.

The measurements for our experiments were made at All-Russian Scientific Research Institute of Experimental Physics (Sarov).

From the standpoint of practical realization of the states of a high-temperature plasma compressed by the emitted gravitational field, two circumstances are of importance.

First. Plasma must comprise two components, with multiply charged ions added to hydrogen, these ions being necessary for quenching spontaneous emission of electrons from the ground energy levels in the own gravitational field. For this purpose it is necessary to have ions with the energy levels of electrons close to the energy levels of free excited electrons. Quenching of the lower excited states of the electrons will be particularly effective in the presence of a resonance between the energy of excited electron and the energy of electron excitation in the ion (in the limit, most favorable case — ionization energy). An increase of z increases also the specific power of the gravitational bremsstrahlung, so that on the condition $\omega_g \leq 0.5\sqrt{10^3 n_e}$ being fulfilled, the equality of the gas-kinetic pressure and the radiation pressure

$$k(n_e\, T_e + n_i\, T_i) = 0.16(0.17\ 10^{-39}\ z^2\ n_e\ n_i\ \sqrt{T_e}\ )\Delta t \tag{30}$$

will take place at $\Delta t = (10^{-6} - 10^{-7})$ s for the permissible parameter values of compressed plasma $n_e = (1 + a)\ n_i = (10^{25} - 10^{26})$ m$^{-3}$, $a > 2$, $T_e \approx T_e = 10^8$ K, $z > 10$.

Second. The necessity of plasma ejection from the region of the magnetic field with the tentative parameters $n_e = (10^{23} - 10^{24})$ m$^{-3}$, $T_e = (10^7 - 10^8)$ K with subsequent energy pumping from the magnetic field region.

The concept of a thermonuclear reactor on the principle of compressing dense high-temperature plasma by emitted gravitational field is supported by the processes of micropinching multicharged ion plasma in pulse high-current diodes. Figs 4, 5 (a) show characteristic parts of micropinch soft X-ray radiation spectrum. Micropinch spectrum line widening does not correspond to existing electromagnetic conceptions but corresponds to such plasma thermodynamic states which can only be obtained with the help of compression by gravitational field, radiation flashes of which takes place during plasma thermalization in a discharge local space. Such statement is based on the comparison of experimental and expected parts of the spectrum shown in Fig. 4. Adjustment of the expected spectrum portion to the experimental one [V.Yu. Politov, Proceedings of International Conference "Zababakhin Scientific Proceedings", 1998] was made by selecting average values of density $\rho$, electron temperature $T_e$ and velocity gradient VU of the substance hydrodynamic motion.

As a mechanism of spectrum lines widening, a Doppler, radiation and impact widening were considered. Such adjustment according to said widening mechanisms does not lead to complete reproduction of the registered part of the micropinch radiation spectrum. This is the evidence (under the condition of independent conformation of the macroscopic parameters adjustment) of additional widening mechanism existence due to electron excited states and corresponding gravitational radiation spectrum part already not having clearly expressed lines because of energy transfer in the spectrum to the long-wave area.

That is to say that the additional mechanism of spectral lines widening of the characteristic electromagnetic radiation of multiple-charge ions (in the conditions of plasma compression by radiated gravitational field) is the only and unequivocal way of quenching electrons excited states at the radiating energy levels of ions and exciting these levels by gravitational radiation at resonance frequencies. Such increase in probability of ion transitions in other states results in additional spectral lines widening of the characteristic radiation. The reason for quick degradation of micropinches in various pulse high-currency discharges with multiple-charge ions is also clear. There is only partial thermolization of accelerated plasma with the power of gravitational radiation not sufficient for maintaining steady states.

The fulfillment of the above-cited conditions (in principle, irrespective of a particular scheme of the apparatus in which these conditions are realized) solves solely the problem of attaining hydrostatic equilibrium states of plasma. The use of a multielectron gas (carbon) as the additive to hydrogen leads to the realization of nuclear fusion reaction conditions, since carbon will simultaneously will act as a catalyst required for the nuclear fusion reaction.

Another variant of nuclear fusion in compositions with multielectron atoms, such as krypton, xenon (and allied elements) is the use of a deuterium-tritium mixture as the light component.

An analysis of the processes which take place in the known devices for generating stable high-temperature states of a plasma (as well as the absence of encouraging results) suggests that the magnetic field can be used only partially, in the first step for the retention and heating plasma in the process of forming its high-energy state. Further presence of the magnetic field no longer confines the plasma within a limited volume, but destroys this plasma owing to the specific character of motion of charged particles in the magnetic field. A principal solution of the problem is a method of confining of an already heated plasma in an emitted gravitational field in a second step, after the plasma has been compressed, heated and retained during this period by the magnetic field. As follows from the above-stated, under any circumstances. Plasma must be injected from the magnetic field region, but with subsequent pumping of energy from the region of the plasma found in the magnetic field. It is just to these conditions that, among other things, there corresponds the original circuit diagram of a magnetoplasma compressor, presented in the specification to the Application.

**The claimed method is realized in the following manner (see the diagram):** through a quick-acting valve 1 a two-component gas (hydrogen + a multielectron gas) is supplied into a gap between coaxial conical electrodes 2, 3, to which voltage is fed through a discharge circuit 5. A discharge creating a magnetic field flows between the electrodes. Under the pressure of the arising amperage, plasma is accelerated along the channel. At the outlet in a region 7 the flow converges to the axis, where a region of compression with high density and temperature originates. The formation of the region of compression 7 is favored by efflux currents which flow in the outgoing plasma jet. With the voltage fed to the anode (20—40) kV and the starting pressure of the working gas (0.5— 0.8) mm Hg, and when the current in the forming discharge reaches about 1 MA in the region of compression, the values of the plasma parameters $n_e$, $n_i = (10^{23}—10^{25})$ m$^{-3}$ and of the temperatures $T_e$, $T_i = (10^7—10^8)$ K, necessary for the excitation of the gravitational field of the plasma, will be reached. The presence of the ions of multielectron atoms in the composition of the working gas, which lead to quenching the gravitational emission from the ground energy levels of the electrons, and cascade transitions along the levels of the electron stationary states in the own gravitational field will lead to the transformation of the high-frequency emission spectrum into the lows-frequency one with frequencies corresponding to locking and amplification of the plasma emission. Simultaneously the density and temperature of the plasma will grow owing to its pulsed injection. Therefore, sub sequent compression of the plasmas after its injection from the magnetic field region to the state of hydrostatic equilibrium (formation of the stable stat of the dense high-temperature plasma) takes place owing to the excitation, locking and compression of the  plasma by the radiated gravitational field, with the attainment of the plasma parameters $n_e$, $n_i = (10^{25}$—$10^{26}$) m$^{-3}$ and $T_e$, $T_i = 10^8$ K

The fundamental difference of such scheme from the known schemes used for obtaining plasmodynamic discharges (Kamrukov A.S. et al., "Generators of laser and powerful thermal radiation, based on heavy-current plasmodynamic discharges" in the book "Plasma Accelerators and Ion Guns", Moscow, Nauka, 1984 (in Russian)), when using this scheme as a quantum generator of gravitational emission (a quantum generator of gravitational emission being just the generator of stable high-energy states of a dense plasma) is as follows:

1. The pulsed character of the discharge circuit with the volt-ampere characteristics corresponding to the plasmodynamic discharge;

2. The definite ratio of the hydrogen component and of the multielectron gas component (approximately 80% and 20%, respectively), including the purposed of attaining the required temperature parameters of the plasma.

3. The close correspondence of the electron energy levels in the employed multielectron gas with the lower electron energy level in the own gravitational field, which requires using such gases as krypton and xenon as the multielectron gas. Here both the percentage of the multielectron atoms, limited from below by the requirement of quenching the excited lower energy states of the electron (in the own gravitational field) and from above by the requirement of attaining the necessary plasma temperature should be adjusted.

The sequence of the operations is carried out in a two-sectional chamber of MAGO installation [Fig. 6, developed at All-Russian Scientific Research Institute of Experimental Physics (Sarov)]; the structure of the installation is most suitable for the claimed method of forming steady states of the dense high-temperature plasma) with magnetodynamic outflow of plasma and further conversion of the plasma bunch energy (in the process of quenching) in  the plasma heat energy for securing both further plasma heating and exciting gravitational radiation and its transit into a long-wave part of the spectrum with consequent plasma compression in the condition of radiation blocking and increasing.

5      The MAGO chamber capacity of work with deuterium-tritium composition was tested experimentally. The obtained experimental data of plasma compression in MAGO chamber [R. Linbemuth et al., Physical Rew. Let., V. 75#10, p.1953, (1995)] prove that there is the fusion reaction (Fig. 7); however the holding time is not sufficient, there need to be longer holding time. The choice of such design as a design for a thermonuclear

10    reactor is unequivocal since it is completely corresponds to the system of exciting and amplifying gravitational radiation when plasma is thermolized after outflow from the nozzle, and required additional compression actually takes place when the working plasma composition is changed (many-electron ions) and current-voltage characteristic of the charge changes correspondingly. The simplicity of the MAGO chamber technical structure

15    is even more clearly shown by the possibility to use as the generator of electrical load such devices as a capacitors battery or an autonomous magnetic explosion generator (VMG) with all consequences of practical use of such thermonuclear reactor.

       One skilled in the art will understand that various modifications and variants of embodying the invention are possible, all of them being comprised in the scope of the

20    Applicant's claims, reflected in the set of claims presented hereinbelow.

**CLAIMS:**

1. A method of forming stable states of a dense high-temperature plasma, which comprises the following steps:

5  a) generation of a dense high-temperature plasma from hydrogen and isotopes thereof with the aid of pulsed heavy-current discharges;

b) injection of the plasma from the area of a magnetic field with parameters corresponding to the conditions of gravitational emission of electrons with a banded energy spectrum;   and

10  c) energy transfer along the spectrum, performed by cascade transition into the long wavelength region of eV-energy to the state of locking and amplification of the gravitational emission in the plasma and simultaneous compression to  the states of hydrostatic equilibrium,

wherein in the formation of the states indicated in step c), in the composition of a

15  working gas multielectron atoms are used for quenching the spontaneous gravitational emission from the ground energy levels of the keV-region electron in the proper gravitational field.

2. A method according to claim 1, in which hydrogen and multielectron atoms are used  for obtaining stable states of a dense high-temperature plasma.

20  3. A method according to claim 1, in which hydrogen and carbon are used for realizing the conditions of the nuclear fusion reaction to proceed, wherein carbon is used both for quenching the spectra of the gravitational emission with keV energies and as a fusion reaction catalyst.

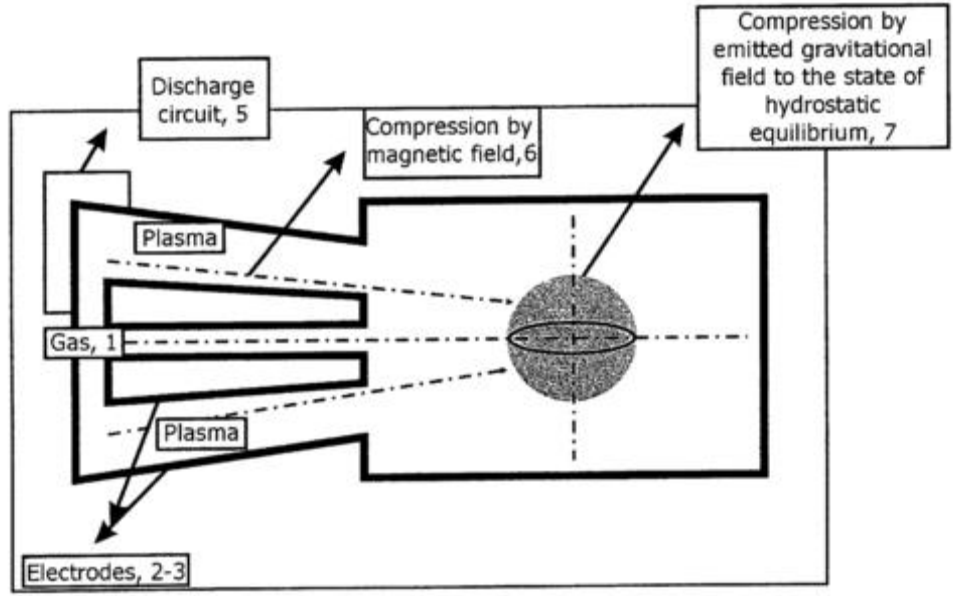

Fig.1

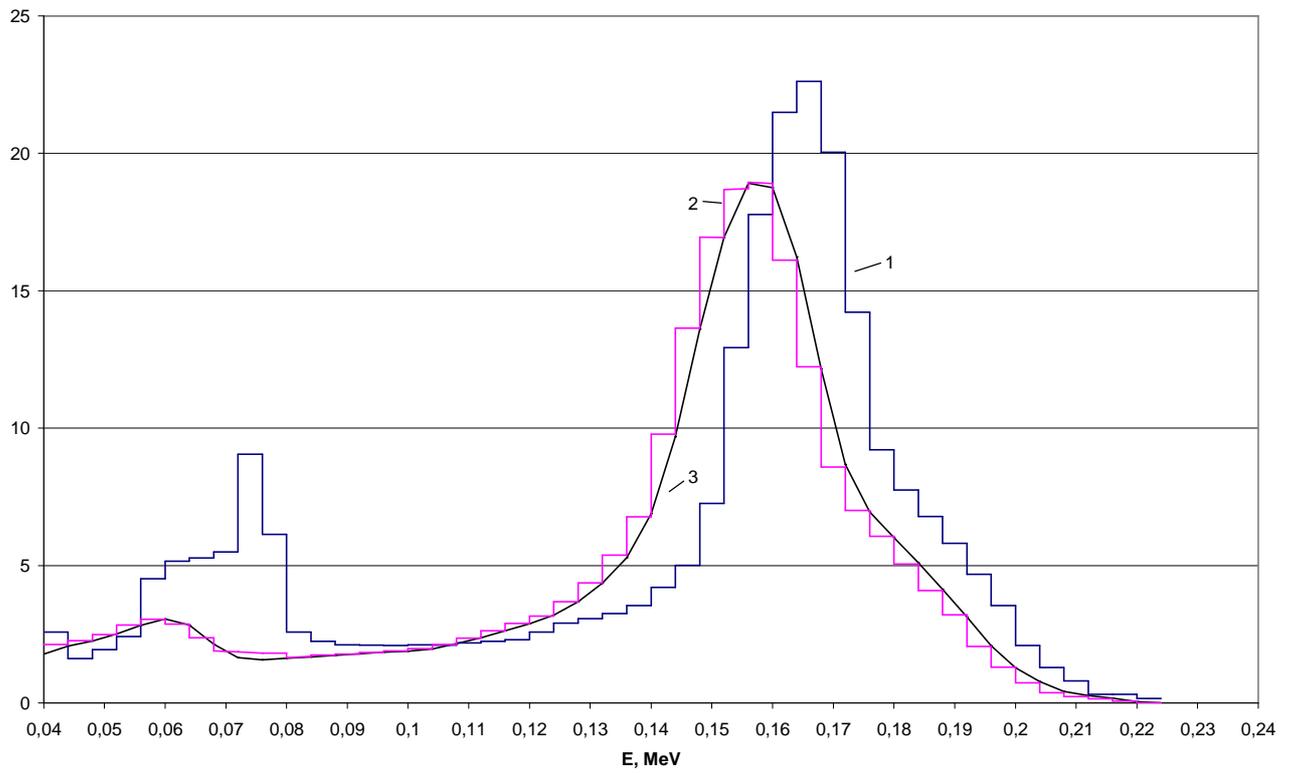

$$\frac{\Delta N_e}{\Delta E}, \frac{электрон}{МэВ}$$

*Figure 2. Electron energy spectra: 1 – after passing the grid, 2 – after passing the Al foil 13 μm thick; 3 – spectrum calculation according to ELIZA program based on the database [D.E. Cullen et al., Report IAEA-NDS-158, September, 1994] for each spectrum 1. The spectrum is normalized by the standard.*

$$\delta\left(\frac{\Delta N_e}{\Delta E}\right), \frac{электрон}{МэВ}$$

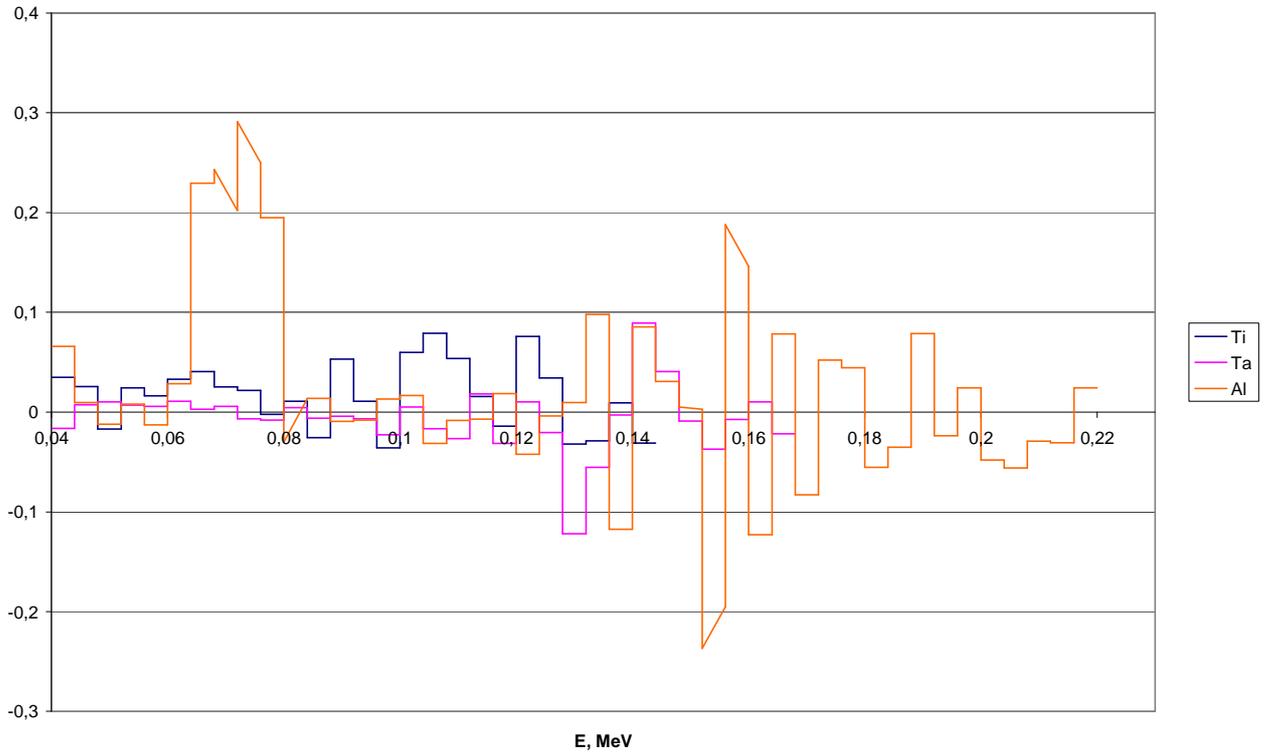

*Figure 3. Difference of spectral density for theoretical and experimental spectra of electrons passed through Ti, Ta and Al foils.*

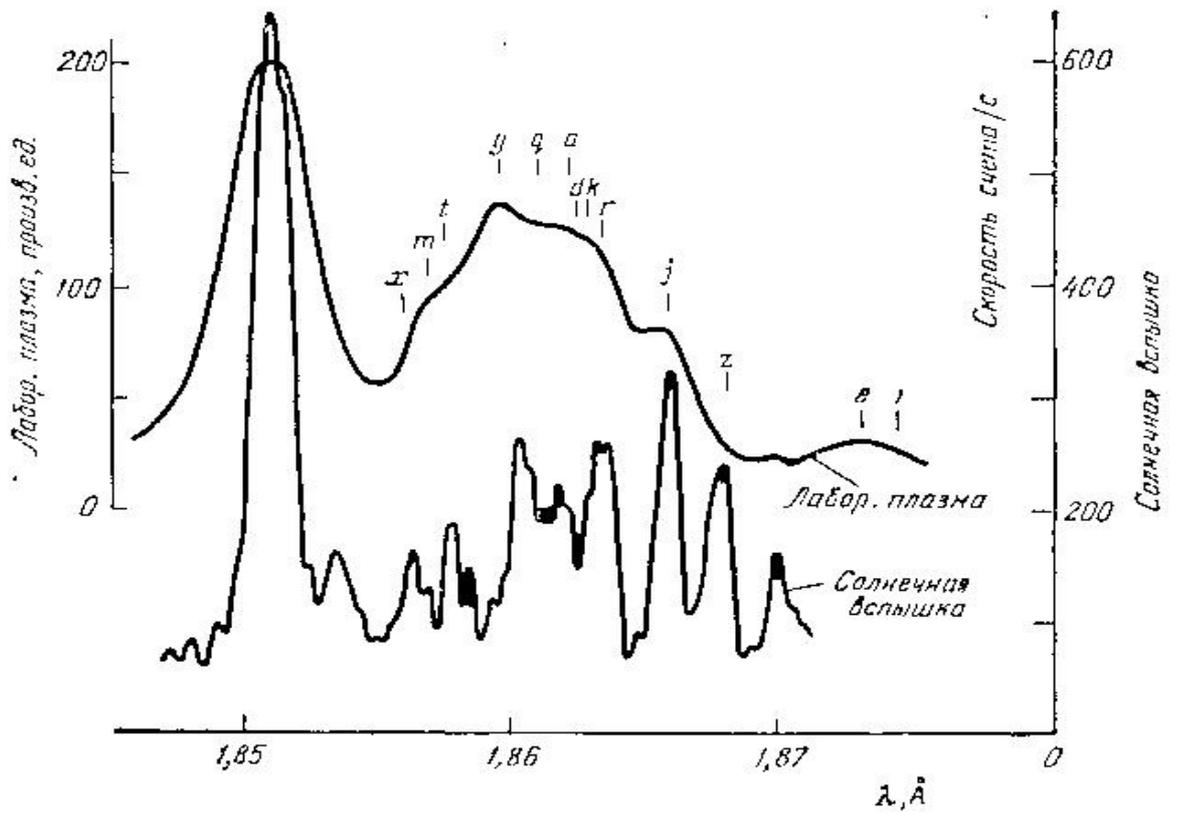

*Figure 4. A part of vacuum sparkle spectrum and a corresponding part of solar flare spectrum. [E.Ya. Goltz et al., DAN USSR, Ser. Phys., 1975, V.220, p.560].*

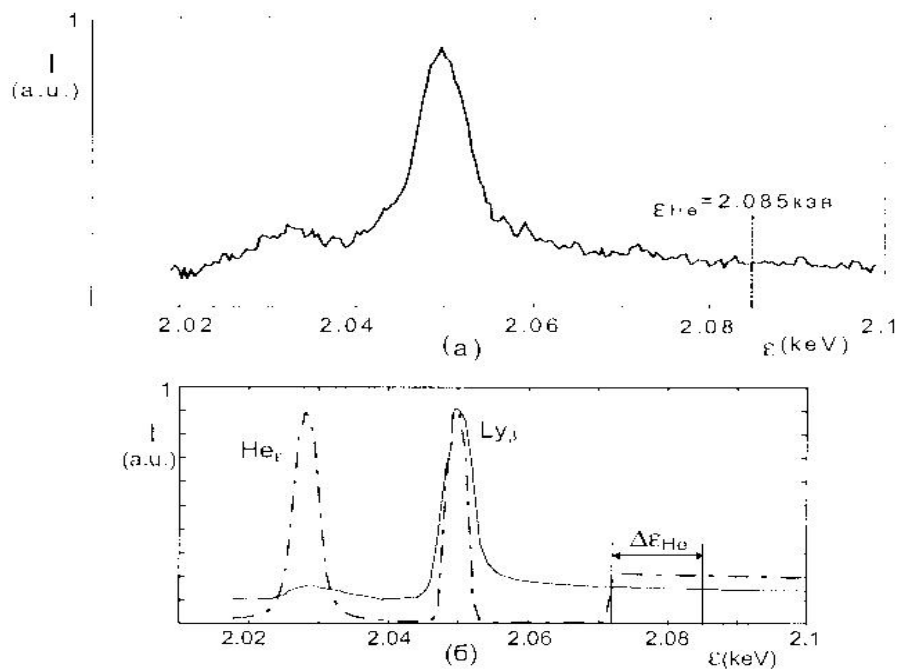

*Figure 5. Experimental (a) and calculated (b) parts of a micropinch spectrum normalized for line Lyβ intensity in the area of the basic state ionization threshold of He-like ions.*

*The firm line in variant (b) corresponds to density of 0.1 g/cm³ , the dotted line – to 0.01 g/cm³; it was assumed that $T_e = 0.35$ keV, [V.Yu. Politov, Proceedings of International Conference "Zababakhin Scientific Proceedings", 1998].*

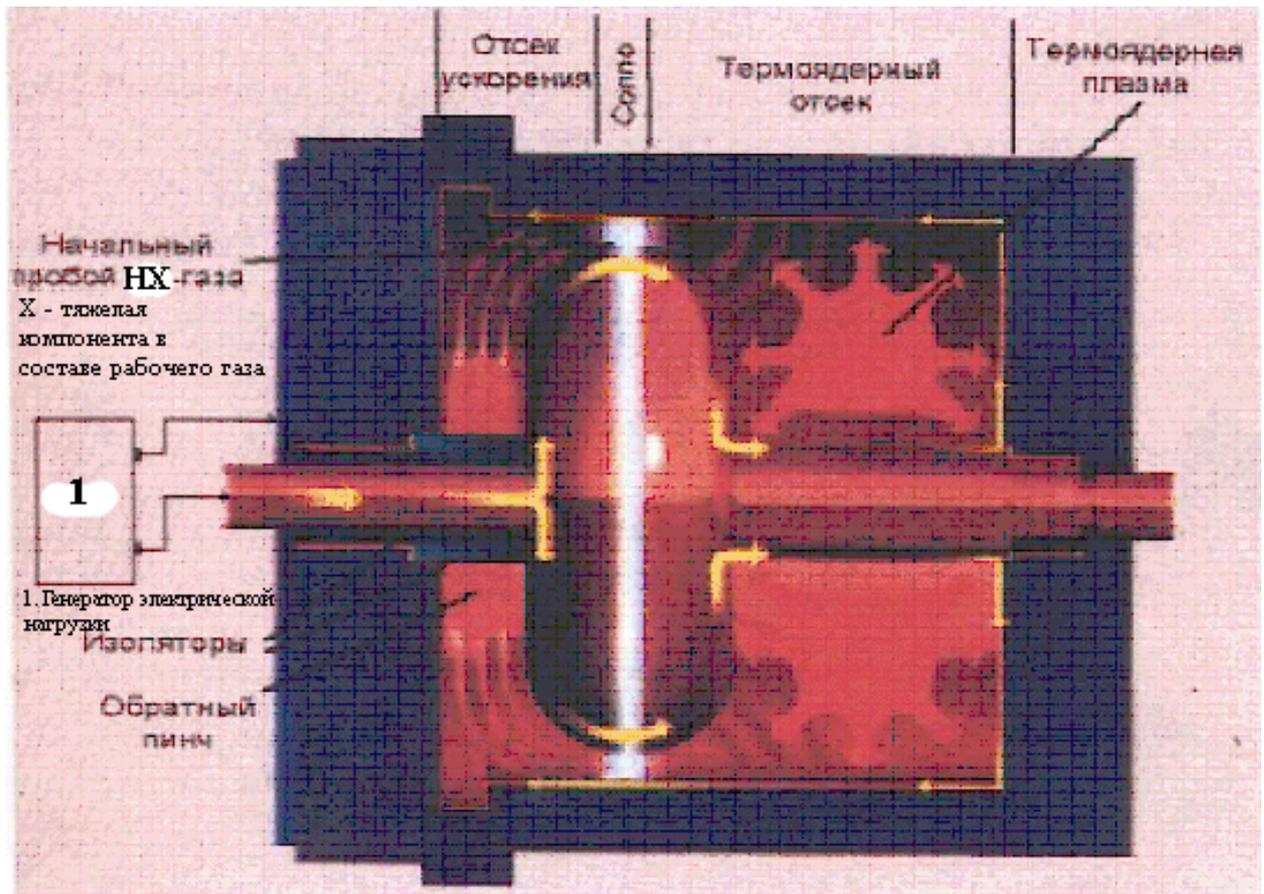

*Figure 6. Physical diagram of thermonuclear plasma in MAGO chamber.*

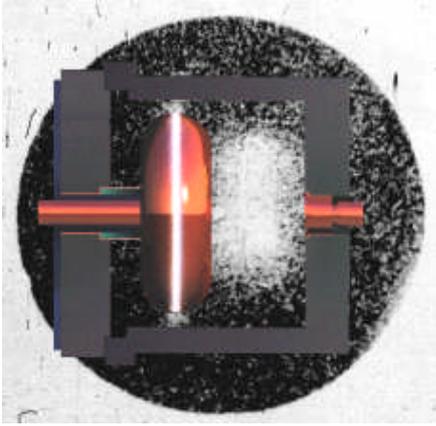

*Figure 7. Neutron picture of the neutron generation area.*